\documentclass[a4paper,twocolumn,11pt,accepted=2022-01-28]{quantumarticle}
\pdfoutput=1
\usepackage{bm, graphicx, amsmath}
\usepackage{bbm}
\usepackage{amssymb,amsmath}
\usepackage[mathscr]{eucal}
\usepackage{graphicx}
\usepackage[draft,inline,nomargin]{fixme} \fxsetup{theme=color}
\usepackage[normalem]{ulem}

\usepackage[numbers,sort&compress]{natbib}

\usepackage{hyperref}
\hypersetup{
colorlinks=true,
linkcolor=red,
urlcolor=magenta,
citecolor=blue
}

\newcommand{\tr}{{\rm tr}\,}
\newcommand{\ket}[1]{\left|{#1}\right\rangle}
\newcommand{\bra}[1]{\left\langle{#1}\right|}
\newcommand{\braket}[2]{\langle{#1}|{#2}\rangle}
\newcommand{\ketbrad}[1]{\left|{#1}\rangle\!\langle{#1}\right|}

\usepackage{tcolorbox}

\newcommand{\be}{\begin{equation}}
\newcommand{\ee}{\end{equation}}
\begin{document}
\title{Online identification of symmetric pure states}
\author{Gael Sent\'{\i}s}
\author{Esteban Mart\'{i}nez-Vargas}
\author{Ramon Mu\~{n}oz-Tapia}
\affiliation{F\'isica Te\`orica: Informaci\'o i Fen\`omens Qu\`antics, Departament de F\'isica, Universitat Aut\`onoma de Barcelona, 08193 Bellatera (Barcelona) Spain}

\begin{abstract}
We consider online strategies for discriminating between symmetric pure states with zero error 
when $n$ copies of the states are provided. Optimized online strategies involve local, possibly 
adaptive measurements on each copy and are optimal at each step, which makes them horizon independent, hence robust in 
front of particle losses or an abrupt termination of the discrimination process. We first 
review previous results on binary minimum and zero error discrimination with local measurements 
that achieve the maximum success probability set by optimizing over global measurements, 
highlighting their online features. We then extend these results to the case of zero error 
identification of three symmetric states with constant overlap. 
We provide optimal online schemes that attain global performance for any $n$ if the state 
overlaps are positive, and for odd $n$ if overlaps have a negative value. 
For arbitrary complex overlaps, we show compelling evidence that online schemes fail to reach 
optimal global performance. The online schemes that we describe only require to store the last 
outcome obtained in a classical memory, and adaptiveness of the measurements reduce to at most 
two changes, regardless of the value of $n$. 
\end{abstract}


\maketitle
\section{Introduction}
The task of discriminating among non-orthogonal quantum states~\cite{Chefles2000,Barnett2009, Bergou2010, Bae2015} 
underlies many prominent 
applications of quantum information sciences.
A basic primitive in quantum communication~\cite{Helstrom1976,Gisin2007}, it also has fundamental implications in quantum key distribution~\cite{Bennett1992,Gisin2002,Acin2006,Renner2008}, in the design of quantum algorithms~\cite{Bacon2005}, and in foundations of quantum theory~\cite{Bae2011,Takagi2019,Oszmaniec2019,Uola2019}. 
Due to the no-cloning theorem~\cite{Wootters1982}, it is not possible to perfectly and deterministically identify which is the state of a given quantum system out of a known finite set of possible ones, unless these are mutually orthogonal. If copies of identically prepared systems in the same unknown state are provided, we may extract more information and increase our chances of identifying it correctly. However, in order to take full advantage of these extra resources, one generally needs to apply a collective quantum measurement on all the provided systems, which requires performing entangling operations and keeping all systems to be measured in a coherent quantum memory. Such collective measurement, once optimized, is guaranteed to yield the best performance in the discrimination task allowed by quantum mechanics, but the necessary requirements to implement it are hardly met in practical situations. 

More experimentally viable (albeit generally sub-optimal) schemes are those that only involve local measurements on each system, thus removing the need of quantum memories and quantum correlations in the measurement apparatus. Such schemes fall under the paradigm of \emph{local operations and classical communication} (LOCC). The question of when can LOCC schemes attain optimal (global) performance in a state discrimination task has been considered in the literature under different angles~\cite{Walgate2000,Virmani2001,ChenDong2001,Chen2002,OptimalConclusJiZh2005,Acin2005,Croke2017,Peres1991,Chitambar2013,Cheng2021}. 

The motivation behind this topic is not only practical, but also foundational: a performance gap between optimal local and global schemes in discriminating separable states is a signature of the phenomenon called ``quantum nonlocality without entanglement''~\cite{Bennet1999}, which has implications in the capacities of quantum channels~\cite{Fuchs2002}, in the ability to hide information to classical observers~\cite{Eggeling2002}, and in distinguishing quantum theory from other generalized probabilistic theories~\cite{Bhattacharya2020}.

In this paper, we take a step further from LOCC and consider \emph{online} strategies for state discrimination, that is, feed-forward local measurement schemes that do not depend on knowing beforehand the number of copies of the states available (the horizon), and are optimal at each step of the process. In contrast to horizon-dependent LOCC, online schemes do not loose optimality if some of the systems are lost or if the procedure stops at an earlier time than planned, thus making them the most desirable schemes for robust realistic implementations. This sort of data processing can be regarded as a self-learning process~\cite{Fischer2000}, and it is the natural procedure in sequential analysis algorithms~\cite{wald1973sequential,MartinezVargas2021}.

When trying to discriminate between two states, it is known that online strategies attain optimal global performance, regardless of whether one considers minimum error discrimination~\cite{Acin2005} or unambiguous identification~\cite{ChenDong2001, OptimalConclusJiZh2005}, the two usual approaches to state discrimination. Discriminating more than two hypotheses is a much harder problem: optimal protocols are only known for certain special cases~\cite{Barnett2001,Bae2015,Bergou2012,DallaPozza2015,Krovi2015,Skotiniotis2018,Sentis2017}, and results on local distinguishability are even more scarce~\cite{Peres1991,Chitambar2013,Chitambar2014,Sentis2019,Nakahira2019}. Here we tackle the problem of unambiguous (zero error) identification of three symmetric pure quantum states with constant (but arbitrary) overlap $c$
when $n$ copies are provided, characterizing for which parameter ranges do online schemes attain global performance. We first rederive the case of binary discrimination, highlighting the online features of the optimal local protocols, and then we extend our formalism to three hypotheses. Specifically, we show that online strategies based on Bayesian updating are globally optimal for any $n$ if 
$c\geq 0$,
and for odd $n$ if 
$c < 0$.
Our analysis straightforwardly extends to the case of tensor products of $n$ trines with constant but different overlaps.
Importantly, the choice of each measurement in these strategies depends only on the last outcome obtained, thus greatly limiting the size of the classical memory required. 
For complex-valued overlaps, we provide strong evidence of a gap between online and global strategies. 

The paper is organized as follows. In Sections~\ref{min-error} and~\ref{sec:unambiguous} we review online binary minimum-error discrimination and unambiguous identification, respectively, and extend these results to non-identical copies of the states. This serves us to set notation and techniques that we use later. Section~\ref{3-states} contains our main results for three symmetric states, and we finish with some conclusions of our analysis.

\section{Two-state minimum error discrimination}
\label{min-error}
Here we briefly review binary discrimination for minimum error~\cite{Brody1996,Acin2005} and its extension to the multi-hypothesis case.  

Any two pure states  can be written w.l.o.g. as 
\begin{align}
\label{ket-2}
\ket{\psi_{0/1}}=\sqrt{\frac{1+c}{2}}\ket{0}\pm\sqrt{\frac{1-c}{2}}\ket{1},
\end{align}
where $\ket{0}$ and $\ket{1}$ is a basis of the space spanned by $\{\ket{\psi_0},\ket{\psi_1}\}$  and  $c=|\braket{\psi_0}{\psi_1}|$.  For later reference it is convenient to view this parametrization as $\ket{\psi_0}=\xi_0 \ket{0}+\xi_1\ket{1}$, where $\ket{0}$ and $\ket{1}$ are the eigenstates of the unitary operation $\displaystyle U=\ketbrad{0}+e^{\frac{2i \pi}{2}}\ketbrad{1}=\ketbrad{0}-\ketbrad{1}$, $\ket{\psi_1}=U\ket{\psi_0}$, and
  $\xi_i=\sqrt{\lambda_i(G)/2}$, $i=0,1$, where  $\lambda_i(G)$ are the eigenvalues of the Gram matrix whose elements are $g_{ij}=\braket{\psi_i}{\psi_j}$.  With this parametrization the operator $\Omega=\sum_k\ketbrad{\psi_k}$, which plays a key role in the  extension of larger sets of  symmetric states (Sec~\ref{3-states}), is diagonal, i.e., $\Omega=2 \;\mbox{diag}\{|\xi_0|^2,|\xi_1|^2\}$. 
  
We assume that the two states can occur with arbitrary a priori probabilities $\eta_0$ and $\eta_1$, respectively.  The aim is to minimize the average error probability $P_e=\eta_0 p(1|\psi_0)+\eta_1 p(0|\psi_1)$, or equivalently maximize the success probability  $P_s=\eta_0 p(0|\psi_0)+\eta_1 p(1|\psi_1)$, where $p(r|\psi_i)$, $r=0,1$, is the probability of making the guess $\ket{\psi_r}$ when the state was $\ket{\psi_i}$. These conditional  probabilities are determined by the measurement  $\mathcal{M}$ performed on the system, which is described mathematically as a positive operator-valued measure (POVM). Here the POVM has only two elements $\mathcal{M}=\{E_0,E_1\}$, with  $E_r\geq0$ and $E_0+E_1=\openone$. The Born rule dictates that $p(r|\psi_i)=\tr [E_r \ketbrad{\psi_i}]$. The  optimal success probability has the well known expression~\cite{Helstrom1976}
\begin{equation}
\label{me-1}
P_s=\frac{1+\sqrt{1-4\eta_0\eta_1 c^2}}{2}.
\end{equation}
It is also well known that this success probability is attained with a POVM with elements that are the projectors on the positive and negative spectrum of the operator $\Gamma=\eta_0\ketbrad{\psi_0} -\eta_1\ketbrad{\psi_1}$, the so-called Helstrom measurement~\cite{Helstrom1976}.

The generalization to the multi-copy case 
is straightforward. The optimal value of the success probability $P_s(n)=\eta_0 p(0|{\psi_0}^{\otimes n})+\eta_1p(1|{\psi_1}^{\otimes n})$ is obtained by simply replacing $c\to c^n$ in Eq.~\eqref{me-1}, i.e.,
\begin{equation}
\label{me-2}
P^G_s(n)=\frac{1+\sqrt{1-4\eta_0\eta_1 c^{2n}}~}{2},
\end{equation}
where the superscript $G$ stands for global. The global measurement attaining this bound acts jointly on the $n$ copies, hence a quantum memory to store the systems is required. Note also that it may involve entangling operations between the systems. 

Let us now succinctly show that there exists a scheme  where each system is measured locally and still achieves the optimal success probability given by Eq.~\eqref{me-2}.   It consists of a sequence of Helstrom measurements on each system where prior probabilities are updated at each step $k$ according to the Bayes rule
\begin{align}
\label{bayes-update-1}
 \eta_i^{(k)} (r_k) =: & \eta_i^{(k)}= p(\psi_i| r_k)  \nonumber \\
=  & \frac{\eta_i^{(k-1)} p(r_k|\ \psi_i)}
{\eta_0^{(k-1)} p(r_k|\psi_0)+\eta_1^{(k-1)}p(r_k|\psi_1)}.
\end{align}
Here $r_k=0,1$ is the outcome value of the $k$'th measurement and we have streamlined the notation when no confusion arises.

The crucial property is that the  Helstrom measurements yield the relation
$\eta_0^{(k)} \eta_1^{(k)}=\eta_0^{(k-1)} \eta_1^{(k-1)} c^2$ for any value of the outcome $r_k$ (see \cite{Acin2005}), and thus $\eta_0^{(k)} \eta_1^{(k)}=\eta_0 \eta_1 c^{2k}$
that, once inserted in  Eq.~\eqref{me-1} for $k=n-1$, precisely gives Eq.~\eqref{me-2}.

The Bayes rule \eqref{bayes-update-1} can be seen as a learning process that updates our belief on the occurrence of each state. Observe that the optimal value of the success probability is obtained at each step.  This is an online procedure as the  knowledge of the total number of systems that are available for measurement is not required, in contrast, e.g., to dynamic programming problems where the knowledge of the horizon is needed to carry out an optimization in reverse~\cite{dynamic}. 
Furthermore, measurements in this local scheme only depend on the previous outcome (as opposed to the whole sequence of previous outcomes), thus the size of the required classical memory is minimal.

Interestingly, the same Bayesian updating local protocol turns out to be optimal in the non-\textit{i.i.d.} case, i.e.,  for two arbitrary multipartite product states $ \ket{\Phi}=\ket{\phi_1}\otimes\ket{\phi_2}\otimes \cdots \otimes \ket{\phi_n}$ and $\ket{\Psi}=\ket{\psi_1}\otimes\ket{\psi_2}\otimes \cdots \otimes \ket{\psi_n}$ with arbitrary priors $\eta_\Phi$ and $\eta_\Psi$, respectively \cite{Brandsen2020a}.  The overlap in this case is $C=|\braket{\Phi}{\Psi}|=c_1c_2\cdots c_n$, with
 $c_k=|\braket{\phi_k}{\psi_k}|$. We proceed as in the \textit{i.i.d.} case, that is,  we  perform a series of local Helstrom  measurements  with sequentially  updated priors
and  get
$\eta_\Phi^{k} \eta_\Psi^{k}=\eta_0 \eta_1 c_1^2 c_2^2\cdots c_k^2$.
The success probability  then reads
\begin{align}
P_s &=\frac{1+\sqrt{1-4 \eta_\Phi\eta_\Psi c_1^2c_2^2\cdots c_n^2}}{2}  ,\nonumber \\
\end{align}
i.e., the optimal success probability, Eq.~\eqref{me-1} with $c \to C$.

Going beyond the binary case is much more involved as there are no closed expressions for the success probability for arbitrary priors. 
Optimal solutions (single or multi-copy) are known only in very few cases, that essentially  correspond to symmetric instances (see e.g.~\cite{Bae2015}).
Notice that in any local protocol, even with symmetric sources, the updating rule will necessarily bias the priors and hence render the problem intractable analytically.
One can nevertheless carry out a numerical study. 
It has been recently shown numerically that  local measurements supplemented with the Bayesian updating rule do not yield the optimal global success probability in the minimum error approach already in the case of  three symmetric states~\cite{Brandsen2020b} (see also~\cite{Nakahira2018} for an analysis with symmetric coherent states).  However, it remains an open question whether this feature also holds for zero-error protocols, which we discuss next. 
\section{Two-state zero-error identification}
\label{sec:unambiguous}
We now turn our attention to protocols that identify a state
without errors at the expense of having inconclusive outcomes, a task also 
known as unambiguous discrimination~\cite{Chefles1998,Chefles2001}. Here we show  that in the 
binary case there is also  
a local online procedure that gives the maximum success probability provided by the 
most general global POVM  acting on all systems.

The zero-error POVM in principle has 
three elements: $F_0$ and $F_1$, that unambiguously detect $\ket{\psi_0}$ and $\ket{\psi_1}$, 
respectively, and $F_I$, which we associate 
to an inconclusive outcome.
In order to achieve optimality the success 
probability $P^u_s=\eta_0 P(0|\psi_0)+\eta_1P(1|\psi_1)=:\eta_0 p_0+\eta_1 p_1$ is 
maximized or, equivalently, the inconclusive probability 
$Q=\eta_0 P(I |\psi_0)+\eta_1 P(I|\psi_1)=:\eta_0 q_0+\eta_1 q_1$ is minimized while 
keeping the condition that no errors are committed, i.e
$P(1|\psi_0)=P(0|\psi_1)=0$. Notice that necessarily 
$F_0 \propto \ketbrad{\psi_1^\perp}$ and $F_1\propto \ketbrad{\psi_0^\perp}$, 
therefore the two proportionality constants are the only free parameters. 
It proves useful to cast the problem as a semidefinite program~\cite{Eldar2003} and use the 
conditional success probabilities  $p_0,p_1$ as the parameters to be optimized. The program
reads~\cite{Sentis2017,Vargas2019}
 \begin{equation}\label{sdp-1}
 \begin{split}  
 \max & \ \  \eta_0 p_0+\eta_1 p_1  \\
 \mbox{s.t.}&  \ \  G-\Gamma \geq 0  \\
 & \ \ \Gamma \geq 0\,,
 \end{split}
 \end{equation}
where recall that $G$ is the Gram matrix whose elements are given by the overlaps 
$g_{ij}=\braket{\psi_i}{\psi_j}$, and $\Gamma$ is a diagonal matrix of the conditional 
success probabilities, $\Gamma=\mbox{diag}\{p_0,p_1\}$. The first constraint 
stems from the POVM condition $\openone -F_0-F_1=F_I\geq 0$. 
We note  that this condition does not depend on the priors, only on $G$.  
This is a general feature that applies to any 
number of hypotheses. In the binary case it yields the interesting uncertainty relation
	\be
	\label{uncertainty}
	q_0  q_1\geq c^2,
	\ee
from which the solution of the SDP~\eqref{sdp-1} follows directly: 
	\be
	\label{q0-q1}
	q_0=c \sqrt{\frac{\eta_1}{\eta_0}}, \  \ \   \ q_1=c \sqrt{\frac{\eta_0}{\eta_1}} 
	\ee
if 
\begin{equation}
\label{ua-cond}
c^2\leq \frac{\eta_0}{\eta_1}\leq\frac{1}{c^2} ,
\end{equation}
and either $q_0=1$ and $q_1=c^2$ if $\eta_0/\eta_1\leq c^2$, or  $q_1=1$ and 
$q_0=c^2$ if  $\eta_0/\eta_1\geq 1/c^2$. In these extremal cases the priors are 
so biased that the optimal measurement discards detecting the state with the lowest 
prior and the POVM changes from having three to two elements. For instance,  
in the case 
$q_0=1$ we only have elements  $F_1$ and $F_I$ with $F_1+F_I=\openone$. The 
symmetric case $\eta_0=\eta_1=1/2$ falls inside the range \eqref{ua-cond} for any 
value of the overlap  and yields the well-known  minimum inconclusive 
probability $Q=c$   
 (see, e.g., \cite{Bergou2010}).

The generalization to arbitrary $n$ amounts to do the change $c\to c^n$ in 
Eq.~\eqref{q0-q1}. Note that this replacement  also widens 
 the range of validity 
of the three outcome POVM
\begin{equation}
\label{ua-cond-n}
c^{2n}\leq \frac{\eta_0}{\eta_1}\leq\frac{1}{c^{2n}}.
\end{equation}
This fact plays an important role when discussing local protocols. The minimum average 
success probability  finally reads (here and thereof we assume w.l.o.g. that 
$\eta_0\leq \eta_1$)
        \be
	\label{Q-ua-n}
	Q(n)=\begin{cases} 2 \sqrt{\eta_0\eta_1}c^n & 
        \mbox{if}   \ \ \   \sqrt{\frac{\eta_0}{\eta_1}}\geq c^n \\
		\eta_0+\eta_1 c^{2n}    & 
        \mbox{if}   \ \ \  \sqrt{\frac{\eta_0}{\eta_1}}\leq c^n
	\end{cases}.
	\ee
	
We  next  show that the optimal performance given by Eq.~\eqref{Q-ua-n} 
can always be attained with local measurements. 
At first glance  this result may seem a bit surprising because, for a given $n$ and the same pair of priors, the  global optimal POVM has three outcomes [i.e., Eq.~\eqref{ua-cond-n} is satisfied], while a local one has only  two [i.e., Eq.~\eqref{ua-cond} is not fulfilled]. This mismatch could lead us to think that a local strategy could not attain global optimal performance. However, we note that upon obtaining an inconclusive outcome in a two element local POVM, the priors get updated in such a way that they become more equilibrated. In fact,  there is a step where the updated priors become sufficiently balanced as to satisfy Eq.~\eqref{ua-cond}.  From there on  local POVMs also have three outcomes.  

The proof of the agreement between the local and global procedures for any $n$ and any initial value of the priors goes as follows.
We have to consider the three different ranges of values where the ratio of the priors may lie:
	\begin{align}
		\label{ranges}
		(i)\ \frac{\eta_0}{\eta_1} \leq c^{2n}, 
		 \ (ii)\ c^{2n} \leq \frac{\eta_0}{\eta_1} \leq c^2, \ 
		 (iii)   \ c^2 \leq \frac{\eta_0}{\eta_1} \leq 1 \, .
	\end{align}	
We start addressing range $(iii)$  (note that the symmetric case of equal priors falls in this range). Here both conditions~\eqref{ua-cond} and ~\eqref{ua-cond-n}  
are satisfied for any $n$, i.e., both global and local POVMs give a non-zero probability of detecting any of the states.
The first local measurement is the optimal one  yielding the
inconclusive  probabilities given by  Eq.~\eqref{q0-q1}.
After this measurement, if we have not been successful,  it is straightforward to see that the priors are updated to 
$\eta^1_0=\eta^1_1=1/2$. 
The next measurement is hence optimized  for equal priors, which gives an inconclusive 
outcome with probability $c$ for both sates. Upon failing we repeat the symmetric 
measurement in all subsequent copies. The overall inconclusive probability of this local strategy then reads
\begin{align}
	Q^L(n)& =\eta_0\Pi_{k=1}^{n}q_0^k+\eta_1\Pi_{k=1}^{n} q_1^k\nonumber \\
	&=\eta_0 c \sqrt{\frac{\eta_1}{\eta_0}}c^{n-1}+\eta_1 c 
    \sqrt{\frac{\eta_0}{\eta_1}}c^{n-1}\nonumber \\
	&=2\sqrt{\eta_0 \eta_1} c^n,
\end{align}
i.e., the optimal value in the first case of Eq.~\eqref{Q-ua-n}.

In the range $(i)$ the priors are so biased that, even for a global measurement, it is not 
worth detecting the state $\ket{\psi_0}$. The local procedure consists of a series of 
measurements $\{F_1=|\psi_0^\perp\rangle\langle\psi_0^\perp|, F_I=\ketbrad{\psi_0}\}$ that either detect 
unambiguously $\ket{\psi_1}$ or fail. In this case we have 
\be
Q^L(n) =\eta_0 \times (1)^n +\eta_1 (c^2)^n=\eta_0 +\eta_1 c^{2n},
\ee
which coincides with the second line of Eq.~\eqref{Q-ua-n}. Note that, for large $n$, the
region $(i)$  is increasingly 
small. We would like to stress that, while all the 
measurements are identical, the updated priors are not. Each time one gets an 
inconclusive result the belief that the state is $\ket{\psi_1}$ diminishes and the 
belief in favor of $\ket{\psi_0}$ increases.  This balances the priors, however not enough 
to be worth testing the state $\ket{\psi_0}$.
Indeed, Bayesian updating gives that, for all $k\leq n-1$, 
\begin{equation}
\label{update-1}
\frac{\eta^{(k)}_0}{\eta^{(k)}_1}=\frac{1}{c^2} \frac{\eta_0^{(k-1)}}{\eta_1^{(k-1)}} \to 
\frac{\eta^{(k)}_0}{\eta^{(k)}_1}=\frac{1}{c^{2k}} \frac{\eta_0}{\eta_1}\leq c^2,
\end{equation}
since  $\eta_0/\eta_1\leq c^{2n}$ in this range. 

The most interesting range is $(ii)$. While the global strategy uses a three outcome POVM, the local strategy starts with a 
fully biased  two outcome measurement (because $\eta_0/\eta_1\leq c^2$). Upon obtaining an 
inconclusive outcome, the priors are updated according to Eq.~\eqref{update-1} and  
get more balanced, i.e.,  our belief that the state is $\ket{\psi_0}$ increases. 
We keep doing the same measurement until a step $k_0$ that yields 
${\eta^{(k_0)}_0}/{\eta^{(k_0)}_1}\geq c^2$. This step is guaranteed to be reached 
before $n$, i.e., $k_0<n$. 
Simply observe that
\begin{equation}
\label{update-q-3}
\frac{\eta^{(k_0)}_0}{\eta^{(k_0)}_1}=\frac{1}{c^{2k_0}}\frac{\eta_0}{\eta_1}\geq c^2\to  \frac{\eta_0}{\eta_1}\geq c^{2(k_0+1)},
\end{equation}
which is always compatible with the initial condition of beginning in range $(ii)$ for some $k_0<n$ (the actual value of $k_0$ depends on the particular ratio $\eta_0/\eta_1$).
Therefore, the protocol consists  in  performing a sequence of fixed two-outcome measurements until the $k_0$ step, when we do a three-outcome measurement for biased priors $\eta_0^{(k_0)}$ and $\eta_1^{(k_0)}$, and continue with a sequence of three-outcome measurements for balanced priors as in region $(iii)$ (of course, for as long as we keep on failing). The probability for $n$ failures is
\begin{align}
Q^L(n) =&\eta_0\left[(1)^{k_0}\times c\sqrt{\frac{\eta_1^{(k_0)}}{\eta_0^{(k_0)}}~} \times c^{n-k_0-1}\right]\nonumber \\
             &+\eta_1\left[(c^2)^{ k_0}\times c\sqrt{\frac{\eta_0^{(k_0)}}{\eta_1^{(k_0)}}~} \times c^{n-k_0-1}\right],
\end{align}
 were we have explicitly displayed the terms corresponding to the three different stages of the procedure. Now, taking into account the expression of the updated priors ratio Eq.~\eqref{update-1}, we get
\begin{equation}
Q^L (n) =\eta_0  \sqrt{\frac{\eta_1}{\eta_0}} c^n +\eta_1  \sqrt{\frac{\eta_0}{\eta_1}} c^n =2\sqrt{\eta_0 \eta_1} c^n,
\end{equation}
which again coincides with the global bound, Eq.~\eqref{Q-ua-n}.

We can summarize the procedure in all three regions by the position $k_0$ of the first 
three-outcome local measurement in the sequence. In region $(iii)$, $k_0=0$ and we already start 
with a three outcome local measurement. In region $(ii)$, $k_0\leq n-1$, i.e., the 
accumulated  balance of the priors given by the inconclusive outcomes induces to 
start a three-outcome measurement at some point before reaching $n$. 
Finally, in region $(i)$, for very biased priors the number of copies is not enough 
to abandon the strategy that only detects one of the states.

As in the minimum error case, this local protocol works also in the non-\textit{i.i.d.} case
of product states. One just needs to take into account that
at each step $k$ we have a different overlap $c_k$ and also a different validity 
range Eq.~\eqref{ua-cond}. The minimum failure probability is simply Eq.~\eqref{Q-ua-n} 
with $c^n$ replaced by  $C=c_1c_2\cdots c_n$.

It is worth emphasizing that the local  procedure  described  yields 
the optimal success probability at each step, regardless of total number of systems 
that are finally available for measurement. Besides not requiring quantum memories,  
the local measurement at any given step depends only  on the outcome of the previous 
measurement, hence the size of the classical memory required is minimal. Furthermore, the measurement 
setting at most changes two times.
\section{Zero-error identification of symmetric multiple hypotheses} \label{3-states}

In this section we extend our results to multi-hypothesis scenarios. 
Rather surprisingly, the performance of online sequential strategies and their 
comparison with the global optimal values for zero-error identification have hardly 
been explored. Although even the simplest case of three symmetric states (TSS)  
is quite a big challenge, as discussed in~\cite{Nakahira2019}, the constraints 
imposed by the zero-error requirement provide more chances to obtain analytical 
results. Here we will mainly focus our attention in the TSS case, and also address 
some straightforward generalizations. 

The problem we address consists in doing a zero-error identification of a set of 
states that have equal prior probabilities $\eta_i=1/3$, $i=0,1,2$, and symmetric 
overlaps 
$\braket{\psi_0}{\psi_1}=\braket{\psi_1}{\psi_2}=\braket{\psi_2}{\psi_0}=c$. 
We first analyze the case of positive  
values of $c$, 
and then we address the negative range. 
We finally consider the sequential performance for complex values of $c$.  

The positive range, 
$0\leq c \leq 1$, can actually be solved for any 
number $r$ of hypotheses as we show below. Note that the  anomaly 
identification problem~\cite{Skotiniotis2018} falls under this case. 
The Gram matrix, $G$, together with the priors encapsulate all the discrimination 
properties of an ensemble, and no explicit form of the states is even needed, although 
the very existence of a valid Gram, i.e., $G\geq 0$, imposes some restrictions on 
the states that can give rise to $G$. For instance, if $0\leq c<1$ the states are necessarily linearly 
independent (a requisite to have zero-error discrimination~\cite{Chefles1998}) and 
therefore the dimension $d$ of the Hilbert space of the states must  be at least 
\mbox{$d\geq r$}. The Gram matrix of a set of three states with equal overlap $c$ reads
\begin{equation}
\label{G-3}
G=\begin{pmatrix}
1 & c & c \\
c & 1 & c\\
c & c & 1
\end{pmatrix}.
\end{equation}
In this symmetric setting the optimal conditional success probabilities must be 
identical, $p_i=p$,  hence the 
SDP~\eqref{sdp-1} reads 
 \begin{equation}
 \label{sdp-2}
 \begin{split}
 \max & \ \  p   \\
 \mbox{s.t.}&  \ \  G- p \openone \geq 0  \\
 & \ \ p \geq 0\, .
 \end{split}
 \end{equation}
This optimization gives the minimum eigenvalue of $G$, 
\be
\label{p}
p=\lambda_{\min}(G)=1-c\,,
\ee
i.e., $q=c$. Note that this solution is the same for any number of symmetric hypotheses. Given $n$ copies of the states, 
the minimum inconclusive probability for any set of symmetric states with constant positive overlaps is  $Q=c^n$.

Next we would like to know if the global performance can also be reached with an
online protocol. This way, no quantum memory would be required and the identification process can be completed at much earlier times without compromising the probability of success~\cite{Vargas2021}. The online protocol consists simply in a local optimal unambiguous measurement 
at each step $k$. One stops as soon as a conclusive outcome is obtained. This protocol can be regarded as a Bayesian updating procedure: if the identification is successful, the priors become 1 for the identified state and zero for the rest of states. If one fails, the updated priors are again symmetric. The proof follows directly from the fact that the inconclusive probability at each step is $c$ and  
$n$ consecutive failures have probability  $c^n$. 

The particular form of the unambiguous POVM that we need depends on the specific source states at hand. We present the TSS case ($r=3$) in detail, but the generalization to an arbitrary number of symmetric source states is straightforward.  As already introduced in Section~\ref{min-error}, the most convenient parametrization is to use the eigenbasis of the unitary $U=\ketbrad{0}+e^{2 i \pi/3}\ketbrad{1}+e^{4 i \pi/3}\ketbrad{2}$, and write the states as 
$\ket{\psi_0}=\xi_0\ket{0}+\xi_1 \ket{1} +\xi_2 \ket{2}$, $\ket{\psi_1}=U\ket{\psi_0}$ and $\ket{\psi_2}=U^2\ket{\psi_0}$. Here  the 
amplitudes $\xi_i$ are related to the eigenvalues of $G$, $\lambda_i$,  through 
\be
\label{xi}
\xi_i=\sqrt{\frac{\lambda_i}{3}}, \ \ i=0,1,2, 
\ee
which is the direct extension of Eq.~\eqref{ket-2}. This parametrization can be regarded as the canonical form of  symmetric states for any 
overlap $c$ (real or complex), and generalizes 
trivially to any number of symmetric states.
It is useful to note that the operator $\Omega=\sum_{k=0}^2 \ketbrad{\psi_k}$ is diagonal in this basis:
\be
\label{diagonal}
\Omega =
 3 \begin{pmatrix}
 |\xi_0|^2 & 0& 0\\
 0& |\xi_1|^2 & 0\\
 0 & 0 & |\xi_2|^2
 \end{pmatrix}
\ee
(this property holds true for any set of three symmetric states, normalized or not).
The specific values of $\xi_i$ are
\be
\xi_0=\sqrt{\frac{1+2 c}{3}}\,, \ \  \xi_1=\xi_2=\sqrt{\frac{1-c}{3}}\,.
\ee
The POVM has elements $F_i=p|\tilde{\phi}_i\rangle\langle\tilde{\phi}_i|$, 
$i=0,1,2$, and $F_I=\openone -\sum_{i=0}^2 F_i$, where $p=1-c$, as given in Eq.~\eqref{p}. The unnormalized states  $|\tilde{\phi}_i\rangle$ satisfy the unambiguous condition $\braket{\tilde{\phi}_i}{\psi_j}=\delta_{ij}$ and 
 are constructed from a state $|\tilde{\phi}_0\rangle$  as  $|\tilde{\phi}_k\rangle=U^k |\tilde{\phi}_0\rangle$. With this parametrization the fiducial state  simply reads
\be
\label{orthogonal-state}
|\tilde{\phi}_0\rangle=\sum_{i=0}^2 \sqrt{\frac{1}{3 \lambda_i}}\ket{i}\,.
\ee

Let us next complete the analysis for 
negative values
of the overlap.
We note that  $G\geq 0$ implies that 
$c \geq -1/2$. In the range $c\in [-1/2,0]$, 
the minimum eigenvalue of 
Eq.~\eqref{sdp-2}  changes to $\lambda_{\min}=1-2|c|$.  For a given number of copies $n$ 
the minimum eigenvalue alternates between $1-2|c|^n$ and  $1-|c|^n$ depending on whether $n$ 
is odd or even, respectively. This means that the minimum inconclusive
probability is
\begin{equation}
	\label{q-3-global}
	Q(n)=\begin{cases}
		2|c|^n & \mbox{if $n$ is odd} \\
		|c|^n    & \mbox{if $n$ is even} 
	\end{cases}.
\end{equation}
Note that indeed $Q(n)$ is a decreasing function of $n$ since  $|c|^{2k} \geq 2|c|^{2k+1} \geq |c|^{2k+2}$ if $|c|\leq 1/2$.

A local protocol based on  fixed unambiguous measurements gives a failure probability $Q^L=2^n |c|^n$, which is away from the optimal value by an exponential factor. 
Given such a large gap, one expects that there exist better local protocols.
The analysis of the extremal value  
$c=-1/2$
gives us the clues on how to proceed. For this value one has $\det G=0$, i.e.,  the three states  are linearly dependent. This means that zero-error identification is not possible \cite{Chefles1998} with only one copy. Of course, given $n>1$ copies, the tensored states  become linearly independent with a global Gram matrix $G>0$. The global inconclusive probability is given by Eq.~\eqref{q-3-global} with 
$c=-1/2$.
It is remarkable that $Q(n)$  is the same for $2k$ and $2k+1$ copies of the state, 
$Q(2k)=Q(2k+1)= 2^{-2k}$, i.e., having an additional copy is of no use (a result already noticed in \cite{Chefles2001}).

Although with only one copy it is impossible to unambiguously identify the state, one can still
gather useful information to be used in the following measurements. In particular,
it is possible to  perform a measurement that is able to exclude one of the states~\cite{caves2002} with 100\% probability. It is easy to see  that a POVM with elements $E_k=\frac{2}{3}|\psi_k^\perp\rangle\langle\psi_k^\perp|$, $k=0,1,2$, does the job, as indeed it constitutes a POVM:  $\sum_{k=0}^2 E_k=\openone$. Then, from the second step onwards, one can  proceed with two-state discrimination measurements as in  Section~\ref{sec:unambiguous} with equal priors. The failure probability then reads
\begin{equation}
	Q^L(n)=\left(\frac{1}{2}\right)^{n-1},
\end{equation}
which coincides with the optimal value for odd $n$, Eq.~\eqref{q-3-global}. Hence, this protocol is 
optimal for any odd number of states. For even $n$ it does not reach global optimality, but we conjecture that also in this case no local protocol can do better than this one.

We can now tackle the whole negative range  
$-1/2<c<0$ with local protocols. The idea is to combine unambiguous identification with the state-excluding measurement that has been the key idea to solve the extremal point $c=-1/2$. 
The unambiguous POVM elements are $F_k=(1-2 |c|)|\tilde{\phi}_{k}\rangle \langle
\tilde{\phi}_{k}| $, $k=0,1,2$, where 
$|\tilde{\phi}_k\rangle$ are given in Eq.~\eqref{orthogonal-state} and  above, and $1-2|c|$ is the minimum eigenvalue of $G$ in this range of $c$. The crucial observation is that  it is possible to construct three additional operators $E_l$ that exclude one of  the states and satisfy $E:=\sum_{l=0}^2 E_l=\openone -\sum_{k=0}^2 F_k=:\openone-F$.
Thus, with the first measurement, either a state is identified with certainty (operators $F_i$)  or a state is excluded also with certainty (operators $E_l$). In other words, either we stop or we continue with a two-state unambiguous measurement 
(with equal priors after their update).
Using Eq.~\eqref{orthogonal-state} and  Eq.~\eqref{diagonal} with the ordering $\lambda_0= \lambda_1=1+|c|$ and $\lambda_2=1-2 |c|$, we have
\be
\label{E}
\openone - F 
= \frac{3|c|}{1+|c|} \begin{pmatrix} 1 & 0 & 0 \\ 0 & 1 & 0\\ 0 &0 &0 \end{pmatrix}.
\ee
The operators 
\begin{align}
E_k=\frac{3 |c|}{1+|c|}\ketbrad{\tilde{\varphi}_k}, 
\end{align}
where  $\ket{\tilde{\varphi}_k}=U^k\ket{\tilde{\varphi}_0}$ and $\ket{\tilde{\varphi}_0}=\ket{0}-\ket{1}$,
 satisfy  the desired conditions 
\begin{equation}
\begin{split}
&\bra{\psi_k}E_k\ket{\psi_k}=0,  \ \ \ k=0,1,2 , \\
& E+F 
=\openone\, .
\end{split}
 \end{equation}
 With  this measurement, the  success probability of unambiguously detecting the state is $1-2|c|$, and hence the probability of excluding one state is $2 |c|$.  The following measurements are binary symmetric  which give an optimal inconclusive probability $|c|$, Eq.~\eqref{q0-q1} with $\eta_0/\eta_1=1$. Therefore, after $n$ measurements the overall inconclusive probability reads
\be
Q^L(n)=2|c|^{n},
\ee
which again coincides with the optimal value Eq.{\eqref{q-3-global} for $n$ odd. 
This result also proves that, for negative values of $c$, 
this protocol is the optimal one among all local procedures when the number of states measured is odd.  For even numbers of states, although we do no have a rigorous proof, there are strong evidences that this is also the case.  A measurement can provide three types of information: (i) exclude two states (unambiguous identification), (ii) exclude one of the states (exclusion) or (iii) update our belief over the  different  states (learning). Naturally (i) is the most valuable information. In a convex combination of POVM elements that achieve (ii) and (iii),
note that the overall failure probability with two copies decreases 
if one puts the maximum weight in the elements leading to (ii).
The POVM $\{F_{0,1,2},E_{0,1,2}\}$ maximizes the contribution to the success probabilities  of (i) and (ii) by construction, hence it is presumably the optimal local measurement for any $n$.

Finally, 
for complex overlaps $c=s e^{i \theta}$,
the eigenvalues of the Gram matrix read 
\be
\label{eigenvalues}
 \lambda_k=1+2s \cos\left[\theta+ \frac{2 k \pi}{3}\right], \ \ k=0,1,2 \, .
 \ee
The minimum eigenvalue is $\lambda_1$ for  $0\leq \theta\leq 2\pi/3$, $\lambda_0$ for  
$2\pi/3\leq \theta\leq 4\pi/3$, and $\lambda_2$ for $4\pi/3\leq \theta\leq 2\pi$. 
The positivity of the Gram matrix  imposes some restrictions on the phase $\theta$ for $s>1/2$. 
The region allowed by the physical restriction $G\geq 0$  is the triangle depicted in Fig.~\ref{figure-draft}.  Note that, by symmetry, values of the overlap $c$ 
differing in a phase of  $2\pi/3$ are equivalent. In particular this holds true for the three lines  with $\theta=0,\,2\pi/3,\,4\pi/3$ and the dashed lines with $\theta=\pi/3,\,\pi,\,5\pi/3$. That is, for values of $c$ 
lying in the ``Mercedes-Benz'' lines  of Fig.~\ref{figure-draft}  a protocol of repeated  unambiguous local measurements  provide the same success probability as gathering all the copies and performing an optimal global measurement, for any $n$. For values in the dashed lines of Fig.~\ref{figure-draft} this is only true for odd $n$. 

For complex values of the overlap 
and for $n=2$ copies, it is possible to find a region with a ``Mitsubishi-logo" shape where a sequence of two local measurements yields the same success probability as the global measurement \cite{Nakahira2019}. However, 
the strategy proposed in~\cite{Nakahira2019} is not online, since it requires knowing the horizon.
Indeed, it sacrifices optimality in the first step (by not putting the maximum possible weight on the POVM elements $F_{0,1,2}$) in order to match global performance at the second step.
We have carried out numerical checks by optimizing over online strategies with local POVMs of the form $\{F_{0,1,2},E_{0,1,2},\openone-F-E\}$. Our results indicate that there is no online protocol yielding the global optimal success probability outside the dark blue and magenta lines of Fig.~\ref{figure-draft}.

\begin{figure}[htbp]
\begin{center}
    \includegraphics[scale=0.75]{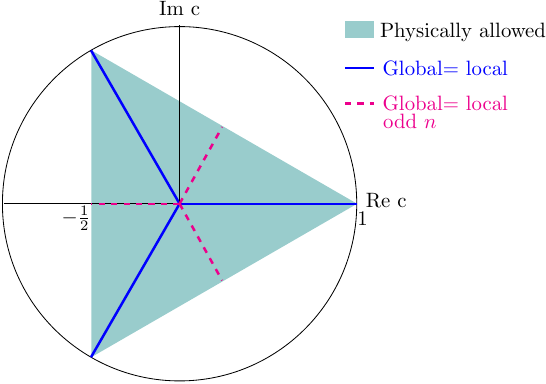} 
\caption{Complex plane of the overlap values. Horizontal and vertical axis correspond to 
    real and imaginary parts, respectively. The shaded triangular region is the 
    physically allowed range. The Mercedes-Benz lines of length one  (solid blue) are 
    the values for which there is an online protocol that matches the optimal performance of 
    global schemes. The rotated lines of length 1/2 (dashed magenta) are the values for which 
    optimality is also attained for odd numbers of copies.}
\label{figure-draft}
\end{center}
\end{figure}

Our results naturally extend to the case
of product states that are not necessarily identical, but
where each local state comes from a different symmetric trine $\{|\psi_0^{(k)}\rangle,|\psi_1^{(k)}\rangle,|\psi_2^{(k)}\rangle\}$ with overlap $c_k$, $k=1,\dots,n$.
This case corresponds to 
a non-\emph{i.i.d.} source that produces three possible global hypotheses
of the form $|\psi_i^{(1)}\rangle |\psi_i^{(2)}\rangle\cdots |\psi_i^{(n)}\rangle$, $i=0,1,2$.
For instance, as in the case of 
identical copies, our online scheme yields the optimal global 
success probability  if 
$c_k\geq 0, \ \ \forall k$. 
Also, if the local trines have positive and negative values of the overlap, the online scheme 
matches optimal performance if $\Pi_{k=1}^n c_k<0$.  
Notice that in this case there must be a first trine with negative overlap, say at step $k$. Recall that the  local measurement for this trine either identifies the state with probability $1-2|c_k|$ or excludes one of the possibilities with probability $2 |c_k|$ and thereafter  one has a symmetric binary problem. Thus, the total inconclusive probability reads  $Q=c_1 c_2\cdots 2|c_k| |c_{k+1}|\cdots|c_n|$ which coincides with the global optimum $2|c_1 c_2\cdots c_n|$  since $c_i>0$ for $i<k$.

\section{Conclusions} 
 
The tasks of binary pure state identification for minimum and zero error can be carried out 
 in an  online fashion with optimal performance. 
The scheme has no horizon, i.e., the information about the number of states available does 
 not affect the measurement scheme.  Optimality is 
 attained at each step regardless of whether systems are lost or one has 
 to stop at an earlier time than planned.  

Extending the analysis beyond the binary case is a much more challenging task. Already the minimum extension of three symmetric states is a highly non-trivial case. For minimum error the direct application of local measurements with Bayesian updating for two copies of the states does not give the optimal global performance~\cite{Nakahira2018,Brandsen2020b}. As far as we are aware, there is no proof that this is the case for more  general one-way local protocols. 

The zero-error identification task, still being  quite involved, offers more 
possibilities to be tackled as 
most of the structure of the POVM is already fixed by the zero-error constraints.  
We have formulated the problem as a semidefinite program that greatly simplifies the 
optimization task and also provides a  very useful tool, not only for numerical 
calculations but, as we exploit here, also for obtaining  analytical results. It also opens 
the path for addressing more complex  instances as, e.g., non-symmetric overlaps or 
different priors. We have given a canonical way of writing  $r$ symmetric states in 
terms of the eigenvalues of their Gram matrix. For $r=3$, we have obtained the optimal online 
protocol for arbitrary positive values of the overlap and any $n$, and for negative values for odd $n$. We have proven that these protocols attain the 
 optimal global performance.
These results directly extend to symmetric complex values of the overlap with phases $2\pi/3$.  
Our findings for positive overlaps also hold for any number of hypotheses $r$.
Unlike~\cite{Nakahira2019}, we are not restricted to sources of linearly independent states. 
We are able to find, e.g., online optimal protocols for trines of symmetric 
qubits. 

For arbitrary complex values of the overlap, our results also suggest that there is no online protocol achieving the same performance as global protocols outside the three symmetric lines of Fig.~\ref{figure-draft}.
The existence of this gap could be exploited in several ways. For instance, one could consider an extension of the B92 protocol~\cite{Bennett1992} with trine states to produce keys of trits. If Alice were to use multiple copies of a trine state for which such gap exists with the objective of increasing the key rate, Bob would take advantage by measuring collectively, while Eve would be forced to measure in an online manner (thus suboptimally) to keep the rate of communication. Another direct application of our results is probabilistic cloning of states from a finite set~\cite{Duan1998} in the asymptotic limit of producing many clones: if the set is a trine where there is no gap between online and global strategies, the task could be optimally performed by an online measure and prepare strategy, thus saving resoures with respect to measuring several copies collectively.
%

\begin{acknowledgments}
We acknowledge discussions with E. Bagan and K. Nakahira.
We acknowledge financial support from the Spanish Agencia Estatal de 
Investigaci\'on, project PID2019-107609GB-I00,  from Secretaria d'Universitats i 
Recerca del Departament d'Empresa i Coneixement de la Generalitat de Catalunya, 
co-funded by the European Union Regional Development Fund within the ERDF 
Operational Program of Catalunya (project QuantumCat, ref. 001-P-001644), and 
Generalitat de Catalunya CIRIT 2017-SGR-1127. EMV thanks financial support from 
CONACYT. 
\end{acknowledgments}
\bibliography{OnlineDis.bib}
\bibliographystyle{quantumrefs}

\end{document}